\documentclass[twocolumn,twoside,slac_two]{revtex4}
\usepackage{graphicx}
\usepackage{fancyhdr}
\begin{document}

\title{Using the EXIST Active Shields for Earth Occultation Observations of X-ray 
Sources}


\author{Colleen A. Wilson, G.J. Fishman}
\affiliation{XD 12, NASA/MSFC, Huntsville, AL 35812}
\author{J.-S. Hong, J.E. Grindlay}
\affiliation{Harvard Smithsonian Center for Astrophysics, Cambridge, MA
02138}
\author{H. Krawczynski}
\affiliation{Washington University in St. Louis, St. Louis, MO 63130}

\begin{abstract}
The EXIST active shields, planned for the main detectors of the coded
aperture telescope, will have approximately 15 times the area of the BATSE
detectors, and they will have a good geometry on the spacecraft for viewing
both the leading and trailing Earth's limb for occultation observations. These
occultation observations will complement the imaging observations of EXIST and
can extend them to higher energies. Earth occultation observations of the hard
X-ray sky with BATSE on the Compton Gamma Ray Observatory developed and
demonstrated the capabilities of large, flat, uncollimated detectors for
applying this observation method. With BATSE, a catalog of 179 X-ray sources was
monitored twice every spacecraft orbit for 9 years at energies above about 25 
keV, resulting in 83 definite detections and 36 possible detections with 
5 $\sigma$ detection sensitivities of 3.5-20 mcrab (20-430 keV) depending on the
sky location. This catalog included four transients discovered with this 
technique and many variable objects (galactic and extragalactic.)  This poster 
describes the Earth occultation technique, summarizes the BATSE occultation 
observations, and compares the basic observational parameters of the 
occultation detector elements of BATSE and EXIST.

\end{abstract}

\maketitle

\thispagestyle{fancy}


\section{Introduction}
\subsection{Earth Occultation Technique}
\begin{figure}
\centering
\includegraphics[width=75mm]{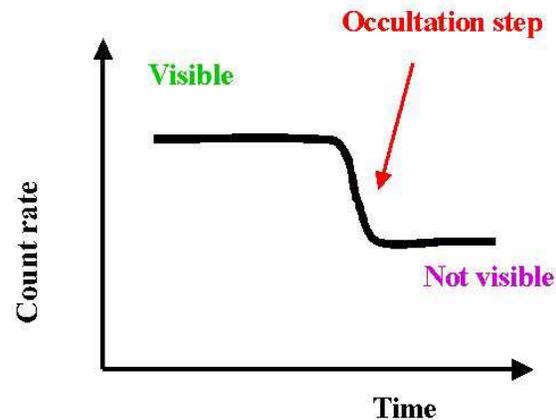}
\caption{Illustration of an Earth Occultation step in count rate data.}
\label{f:step}
\end{figure}

The Earth Occultation technique is simply measuring the change in total count 
rate in a non-imaging detector when an X-ray source rises above or sets below 
the Earth's horizon, due to the spacecraft orbital motion. This produces a 
step-like feature in the data, illustrated in Figure~\ref{f:step}. This 
technique, described in detail in Harmon et al. (2002)\cite{Harmon02}, was 
used very successfully with the Burst and Transient Source Experiment (BATSE) on
the Compton Gamma Ray Observatory to monitor a catalog of 179 sources, of which
83 where definitely detected. This catalog included 32 stellar mass black hole 
systems of which 23 were definitely detected, including 3 newly discovered 
systems, and 12 AGN sources of which 6 were definitely detected. Details of the
BATSE catalog can be found in Harmon et al. (2004)\cite{Harmon04}, including 
data on the web at http://gammaray.msfc.nasa.gov/batse/occultation.

\subsection{EXIST}

\begin{figure*}
\centering
\includegraphics[width=135mm]{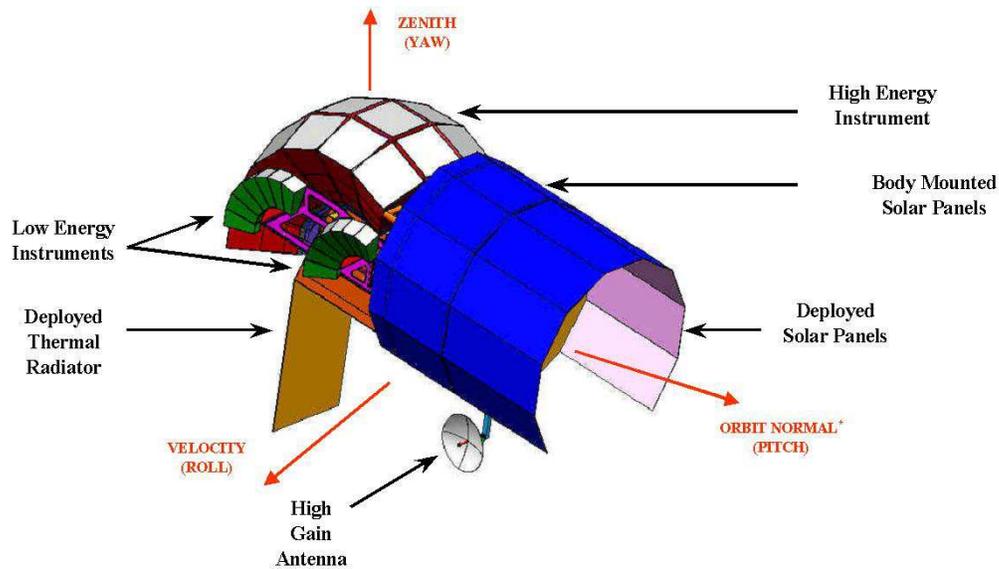}
\caption{Current design concept for the EXIST satellite.}
\label{f:exist}
\end{figure*}

\begin{figure}
\centering
\includegraphics[width=75mm]{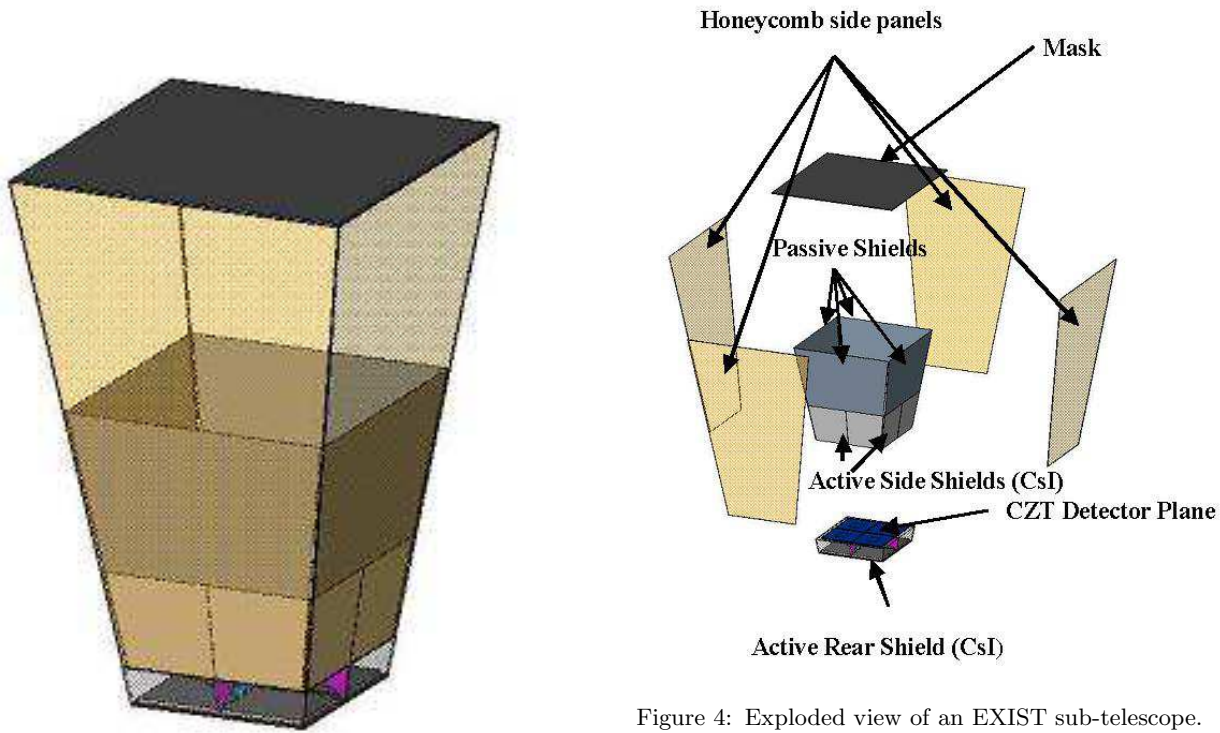}
\caption{One of 18 sub-telescopes in the EXIST high-energy instrument.}
\label{f:subt}
\end{figure}

\begin{figure}
\centering
\includegraphics[width=75mm]{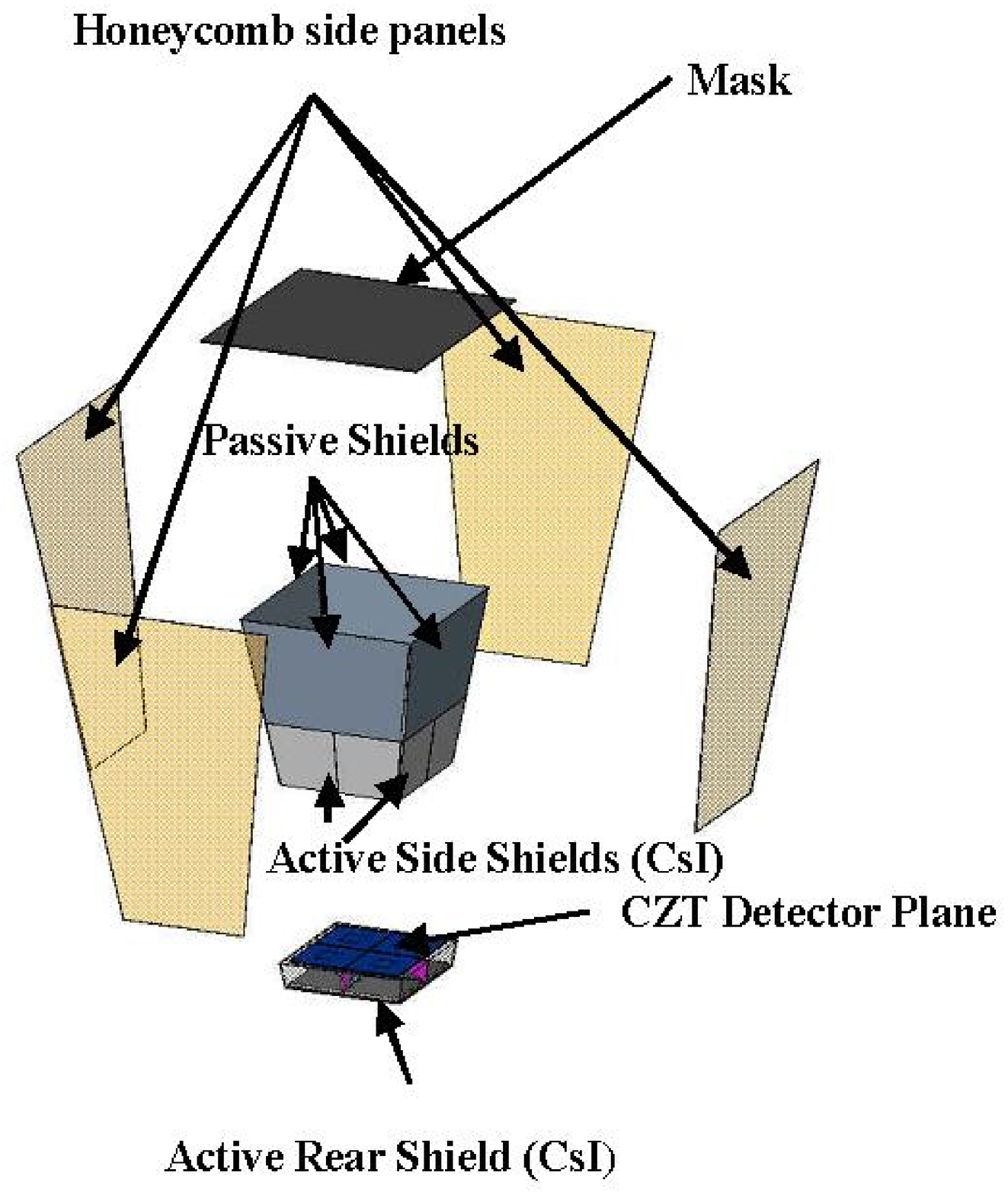}
\caption{Exploded view of an EXIST sub-telescope.}
\label{f:exp_subt}
\end{figure}

The Energetic X-ray Imaging Survey Telescope (EXIST) is a mission concept
for the Beyond Einstein Black Hole finder probe \cite{Grindlay03}. The current
design concept is shown in Figure~\ref{f:exist}. Its primary 
mission will be to study obscured active galactic nuclei and gamma ray bursts. 
The main instrument, the high energy instrument, is a coded mask telescope with
$\sim 6$ m$^2$ of cadmium-zinc telluride (CZT) detectors in the focal plane. It
will be sensitive to photons in the 10-300 keV range. In this poster we propose
using the CsI side shields as Earth occultation detectors to supplement the 
energy range and time coverage of the main instrument. The high energy 
instrument on EXIST will consist of a $6 \times 3$ array of sub-telescopes, one
of which is shown in Figure~\ref{f:subt}. Honeycomb side panels and both active
and passive shields will be shared by adjacent sub-telescopes. The thicknesses
and areas of these active and passive shields will be optimized to meet the
mission constraints. See Figure~\ref{f:exp_subt} for an exploded view of the
current configuration. Table~\ref{t1} lists various parameters for the current 
EXIST concept and BATSE for comparison.

\begin{table}
\begin{center}
\caption{Comparison of EXIST and BATSE}
\begin{tabular}{|l|c|c|}
\hline \textbf{Parameter} & \textbf{EXIST} & \textbf{BATSE} \\
\hline Area & 23100 cm$^2$ & 2025 cm$^2$ \\
            & (array of 6 shields) & (1 BATSE LAD) \\  
\hline Material & CsI & NaI \\
\hline Thickness & 0.8 cm per layer & 1.27 cm \\
                 & (up to 4 layers) & \\
            
\hline Density & 4.51 g cm$^{-3}$ & 3.67 g cm$^{-3}$ \\
\hline Orbit Altitude & 500 km & 400-600 km \\
\hline Orbit Inclination & 7$^{\circ}$ & 28.5$^{\circ}$\\
\hline
\end{tabular}
\label{t1}
\end{center}
\end{table}

\section{Estimating the Earth Occultation Sensitivity of the EXIST Side Shields}

\begin{figure*}
\centering
\includegraphics[width=135mm]{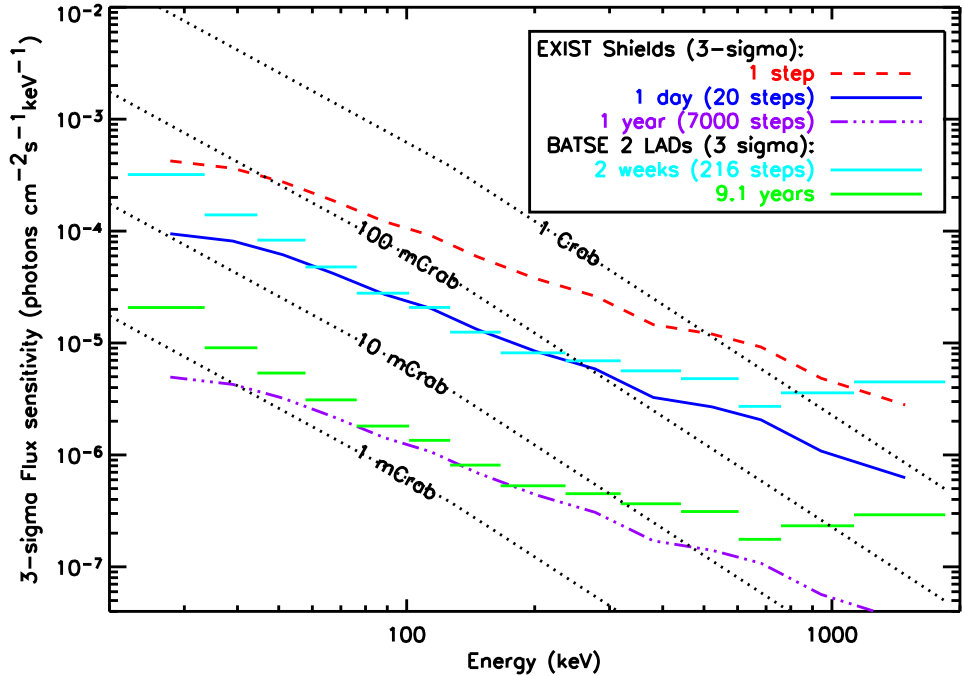}
\caption{Comparison of estimated sensitivity of the EXIST CsI side shields to
the BATSE Large Area detectors for Earth Occultation measurements. Dashed, solid,
and dash-dot curves show EXIST CsI sensitivities for 1 step, 1 day, and 1 year,
respectively. Horizontal bars denote BATSE sensitivity for 2 LADs for 2 weeks 
and for the full 9.1 year mission}
\label{f1}
\end{figure*}

To estimate the sensitivity of the EXIST CsI side shields for Earth Occultation
measurements, we scaled the BATSE large area detector (LAD) count rates from a 
typical interval outside of South Atlantic Anomaly (SAA) passage by the ratio of
the CsI side shield area (2.31 m$^2$) to the area of 1 BATSE LAD (0.2025 m$^2$).
Next we generated a time series of random background rates, using the 16 BATSE 
energy channels. 

We then fit these rates with the following model:                        
\begin{equation}
R(t)  = FA\varepsilon T(t) cos(\theta) + \sum_{i=0}^2 c_i(t-t_0)^i,
\end{equation}
where $F$ is the source flux in photons cm$^{-2}$ s$^{-1}$; $\theta$ is the 
angle to the source; $A$ is the geometric detector area; $\varepsilon$ is the 
detector efficiency; $T(t)$ is the atmospheric transmission function;  $c_i$
are coefficients of a simple quadratic fitted to the background
rate; and $t_0$ is the occultation time of the source.  The atmospheric
transmission function, $T(t)$, depends on the mass attenuation of gamma rays in
air, and the air mass along the line of sight at a given altitude, which in
turn depends upon the precise position of the spacecraft. For EXIST we simply
used an atmospheric transmission function from BATSE. 

The assumed EXIST CsI shield detector efficiency was given by  
\begin{equation}
\varepsilon = (1-e^{-\mu_{\rm CsI}(E) \rho_{\rm CsI} h_{\rm CsI}})
e^{-\mu_{\rm H}(E) \rho_{\rm H} h_{\rm H}} 
\end{equation} 
where $\mu_{\rm CsI}(E)$ and $\mu_{\rm H}(E)$ are the energy dependent mass 
attenuation coefficients for CsI and the honeycomb panels, $\rho_{\rm CsI}$ and
$\rho_{\rm H}$ are the densities of CsI and the honeycomb panels, and 
$h_{\rm CsI}$ and $h_{\rm H}$ are the thickness of CsI and honeycomb panels. We
used 1 CsI layer ($h_{\rm CsI}=0.8$ cm) and 1 honeycomb panel ($h_{\rm H}=0.8$ 
cm) for energies up to about 200 keV. Above 200 keV, we assumed
all 4 CsI layers were used ($h_{\rm CsI} = 3.2$ cm) and that the photons passed 
through 7 honeycomb layers ($h_{\rm H} = 5.6$ cm).

The sensitivity is then given by $N\sigma_{\rm F}$, where $\sigma_{\rm F}$ is 
the computed error on the source flux $F$. In our case we chose N=3 to obtain 
3 $\sigma$ sensitivity estimates, shown in Figure~\ref{f1}.

\section{Conclusions}

Our estimates show that we should reach flux levels in about a day with EXIST
that took an entire 2 week pointing with BATSE and similarly, in about 1 year
we should reach levels comparable or better than those obtained over the entire
BATSE mission. We used a very rough approximation for the background in the
EXIST CsI side shields, obtained by simply scaling the average background in a
BATSE LAD by area, to estimate the Earth Occultation sensitivity for EXIST. One
should keep in mind that our estimates especially at the low and high ends will
be affected by materials in close proximity to the  detector. At low energies,
we have assumed an 8 mm thick honeycomb panel with a density of 0.5 g cm$^{-3}$
sandwiching each shield. At high energies, we simply scaled the BATSE rates,
assuming that the spacecraft materials are similar for BATSE and EXIST. The
background levels at high energies are due to cosmic rays interacting in
materials near the detectors. More detailed background estimates are needed to
better determine Earth Occultation sensitivity for EXIST.

The CZT coded mask imaging system on EXIST will provide more sensitive
measurements of sources especially at energies below 300 keV. Our Earth
Occultation measurements will complement these measurements, providing
additional time coverage when sources are outside the field of view of the main
instrument. This may prove important especially for rapid or short lived
events, such as the rise of a transient stellar mass black hole outburst. In
addition this technique will complement the energy range of the CZT detectors,
providing coverage at energies where the CZT detectors have low efficiency.
\bigskip % extra skip inserted

\end{document}